# Physical properties of the new Uranium ternary compounds U$_3$Bi$_4$M$_3$ (M=Ni, Rh)

T. Klimczuk[1,2], Han-oh Lee[1], F. Ronning[1], T. Durakiewicz[1], N. Kurita[1], H. Volz[1], E. D. Bauer[1], T. McQueen[3], R. Movshovich[1], R.J. Cava[3] and J.D. Thompson[1]

[1] Condensed Matter and Thermal Physics, Los Alamos National Laboratory,

Los Alamos, New Mexico 87545, USA

[2] Faculty of Applied Physics and Mathematics, Gdansk University of Technology, Narutowicza 11/12, 80-952 Gdansk, Poland,

[3] Department of Chemistry, Princeton University, Princeton NJ 08544

**Abstract**

We report the properties of two new isostructural compounds, U$_3$Bi$_4$Ni$_3$ and U$_3$Bi$_4$Rh$_3$. The first of these compounds is non-metallic, and the second is a nearly ferromagnetic metal, both as anticipated from their electron count relative to other U-based members of the larger '3-4-3' family. For U$_3$Bi$_4$Rh$_3$, a logarithmic increase of C/$T$ below 3 K, a resistivity proportional to $T^{4/3}$, and the recovery of Fermi-liquid behavior in both properties with applied fields greater than 3T, suggest that U$_3$Bi$_4$Rh$_3$ may be a new example of a material displaying ferromagnetic quantum criticality.

## 1. Introduction

The hybridization of conduction electrons with more localized *f*-electrons is responsible for the remarkably large quasiparticle masses characteristic of heavy fermion materials. One example of how this hybridization can alter the physical properties of a material occurs in Kondo insulators, where the hybridization creates a gap in the electronic density of states and band filling turns a compound otherwise expected to be metallic into an insulator or semiconductor. One of the best known examples of such behavior occurs in $Ce_3Bi_4Pt_3$ [1]. This structure type is also known for compounds based on uranium, and $U_3X_4M_3$ (X=Sb,Sn; M=Ni,Cu) form, varying between metallic and semiconducting behavior as discussed below. Though this is suggestive of Kondo insulating behavior, the fact that some nonmagnetic Th analogs also display a non-metallic ground state suggests that hybridization may not be responsible for the electronic gap in some of the uranium counterparts.

Physical properties can be tuned by changing the number of electrons in a system. An example is how the superconducting transition temperature, for pure elements and for compounds with the A15 structure, strongly depends on the number of valence electrons[2]. An analogy to this universal rule also works in the uranium "3-4-3" family. For example, $U_3Sb_4Ni_3$ is a semiconductor, and replacing Ni by Cu, which has one more *d*-electron, makes $U_3Sb_4Cu_3$ metallic. Another interesting observation is that both $U_3Sb_4Co_3$ and $U_3Sb_4Cu_3$, which differ by six d-electrons, are ferromagnets with $T_C$ = 10 K and 88 K, respectively [3].

The crystal structure of these 3-4-3 compounds can be understood as a variant of $U_3Sb_4$ with interstices filled by transition metals (M=Ni, Co, Cu, Rh, Pd, Pt, Au), three per formula unit. This stuffing does not change the space group (*I4-3d*) but slightly increases the lattice parameter for example from a=9.113Å ($U_3Sb_4$) [4] to a=9.284 Å for $U_3Sb_4Co_3$ and a=9.684 Å for $U_3Sb_4Pd_3$. Up to now, only the metalloids Sb and Sn have been known to form the $U_3X_4M_3$ structure, and within this family, the electron count, not unit-cell volume, appears to be the dominant factor governing the ground state. Here we show that also Bi can also stabilize this structure, and two new U ternary compounds $U_3Bi_4Ni_3$ and $U_3Bi_4Rh_3$ have been synthesized. Assuming electron count is an indicator of the ground state, we expect $U_3Bi_4Ni_3$ to be non-metallic (as an analog of $U_3Sb_4Ni_3$) and $U_3Bi_4Rh_3$ to be a metallic ferromagnet (analogous to $U_3Sb_4Co_3$). Experiments show that $U_3Bi_4Ni_3$ is non-metallic, possibly due to the appearance of a hybridization gap, and $U_3Bi_4Rh_3$ is a nearly ferromagnetic metal with a logarithmically diverging C/T (specific heat divided by temperature) and low-temperature resistivity that increases as $T^{4/3}$ in zero field. Application of a field to $U_3Bi_4Rh_3$ recovers Fermi-liquid behavior in both specific heat and resistivity, suggesting that $U_3Bi_4Rh_3$ is a new example of ferromagnetic quantum criticality.

## 2. Sample preparation and characterization

Single crystals of $U_3Bi_4M_3$ (M=Ni or Rh) were grown from Bi flux. The pure elements were placed in the ratio 1:10:2 (U : Bi : M) in an alumina crucible and sealed under vacuum in a quartz tube. The tubes were heated to 1150°C and kept at that temperature for four hours, then cooled at the rate of 5 °C /hr to 650°C, at which temperature excess Bi flux was removed in a centrifuge. The resulting crystals were irregularly shaped with typical dimensions 3×3×2mm³. The excess of transition metal (U:M ratio is 1:2) is critical; no crystals were obtained for a starting composition 1:10:1 (U : Bi : M).

$U_3Bi_4Ni_3$ and $U_3Bi_4Rh_3$ crystals were crushed and ground and characterized by powder x-ray diffraction analysis, employing a Bruker D8 diffractometer with Cu Kα radiation and a graphite-diffracted beam monochromator. The software TOPAS 2.1 (Bruker AXS) was used for Rietveld structure refinements. The known crystal structure of $U_3Sb_4Ni_3$ was employed as a starting structural model [5]. Magnetization measurements were performed in a Quantum Design MPMS system. Resistivity and specific heat were measured in Quantum Design PPMS system with a $^3$He insert, using a standard four-probe technique and relaxation method, respectively. Specific heat at low temperature (0.1K ≤ T ≤ 3K and $\mu_0H$ = 0 T) of the $U_3Bi_4Rh_3$ crystal was measured in a $^3$He / $^4$He dilution refrigerator. For resistivity measurements, four platinum wires were attached with silver paint on mechanically cleaved crystal surfaces without polishing or heat treatment, due to the slight air- and heat-sensitive character of these compounds.

For photoemission measurements, two $U_3Bi_4Ni_3$ and two $U_3Bi_4Rh_3$ samples were mounted on a transfer arm, baked at 380 K for 12 hours and transferred into the measurement chamber. Measurements were performed on a SPECS Phoibos 150 electron-energy analyzer working in angle-integrated mode, with an energy resolution of 20 meV. The ultimate resolution of the analyzer (better than 5 meV) was not achieved due to cleave-related irregularities on the sample surface. A helium lamp was used as the excitation source (21.2 eV line). Samples were fresh-cleaved at 15 K in a vacuum of $8*10^{-11}$ Torr.

## 3. Results

An example of the observed x-ray spectra, the calculated powder-diffraction pattern, the difference between the calculated model and experimental data, and positions of expected peaks is presented in Fig. 1 for $U_3Bi_4Ni_3$. The lower set of peaks shows the positions of elemental Bi, which is often present on the crystal surface in the form of small dots. Both compounds were found to be isostructural, with the cubic $Y_3Sb_4Au_3$ - type structure which has a cell parameter of 9.818(1) Å and space group *I4-3d* . As shown in Fig. 1, there is good agreement between the model and the data, and crystal structure of our crystals was confirmed. The lattice parameter for $U_3Bi_4Ni_3$ was calculated to be a = 9.5793(1) Å which is larger than for $U_3Sb_4Ni_3$ (a = 9.393 Å) [6]. Similarly, the lattice parameter for $U_3Bi_4Rh_3$ is a = 9.7273(1) Å, which again is larger than a = 9.501(1) Å for $U_3Sb_4Rh_3$ [7]. These differences stem from the larger covalent radius of Bi and Rh, compared to Sb and Ni, respectively. The refined structural parameters for the new compounds are presented in Table 1.

The electrical resistivity of $U_3Bi_4Ni_3$ (upper panel) and $U_3Bi_4Rh_3$ (lower panel) is plotted as a function of temperature in Fig. 2. These data show that $U_3Bi_4Ni_3$ is non-metallic, as expected by electron count; whereas, $U_3Bi_4Rh_3$, with nominally three fewer electrons, exhibits a positive $\partial\rho/\partial T$, typical of a metal, but with an overall high resistivity that reaches a maximum near 220 K. Conclusions from resistivity are supported by photoemission measurements (Fig. 3) that show a gap in density of states at Fermi level in $U_3Bi_4Ni_3$ and no gap but a reduced density of states in $U_3Bi_4Rh_3$. In order to roughly estimate the gap size form photoemission data, we assume the gap symmetry with respect to zero energy. Lorentzian lineshape is then fitted to a symmetrized density of states, and the electronic gap estimate in $U_3Bi_4Ni_3$ is ≈72 meV. A similarly large gap, ≈95 meV, is deduced from an Arhenius plot of the resistivity for 220 K< $T$ <300 K. Though non-metallic behavior was expected for $U_3Bi_4Ni_3$, the origin of its gap is not obvious. As shown in the inset of Fig. 2, the resistivity of $Th_3Bi_4Ni_3$ also is non-metallic, which superficially suggests that both compounds are not metals because of simple band structure. On the other hand, the valence state of Th is 4+, whereas, susceptibility measurements discussed below are consistent with a U valence state of 3+, 4+ or a value intermediate between these limits. In this case, the electron count in $U_3Bi_4Ni_3$ is similar to that of $Ce_3Bi_4Pt_3$, whose Ce valence is somewhat greater that 3+ and which is semiconducting due to *f*-ligand hybridization.

Magnetic susceptibility χ, measured between 2 K and 350 K under an applied field of 0.1 T, is given in Fig. 4. Above 200 K, the susceptibility of both compounds follows a Curie-Weiss form, and fitting parameters are given in Table 2. The calculated effective magnetic moments are 3.56 $\mu_B$/U-mol and 3.44 $\mu_B$/U-mol for $U_3Bi_4Rh_3$ and $U_3Bi_4Ni_3$, respectively. As mentioned earlier, these values are expected for *5f$^2$* or *5f$^3$* U configurations and are in good agreement with the effective moment obtained for $U_3Sb_4Ni_3$ (3.65 $\mu_B$/U-mol), and slightly higher than found for $U_3Sb_4Rh_3$ (3.2 $\mu_B$/U-mol) . In both compounds, a negative Weiss temperature suggests the presence of antiferromagnetic correlations. At low temperatures, however, the susceptibilities of these materials are very different. Below ~ 60 K, the susceptibility of $U_3Bi_4Ni_3$ rolls over to a nearly temperature-independent value of ~$7 \times 10^{-3}$ emu/mole-U. One possible interpretation of the temperature-independence is that it is due to the Kondo effect, which would give $\chi(0) \approx C/3T_K$, where C is the Curie constant and $T_K$ is the Kondo temperature. Using the high temperature value of C, this relation gives $T_K \approx$ 80 K. Such an interpretation relies on this material being a metal, which it is not. An alternative possibility is that the loss of moment below 60 K reflects the development of a hybridization-induced gap in the spin-excitation spectrum, as found in Ce$_3$Bi$_4$Pt$_3$ [8]. This should be detected in planned neutron-scattering measurements. In contrast to U$_3$Bi$_4$Ni$_3$, there is no evidence for saturation of the susceptibility of U$_3$Sb$_4$Rh$_3$ at low temperatures, and, as plotted in the inset of Fig. 4, the inverse magnetic susceptibility of U$_3$Bi$_4$Rh$_3$ below ~ 4.5 K shows an unusual power-law dependence on temperature $\chi^{-1} \propto T^{\alpha}$, with the exponent $\alpha \approx 0.75$. This power-law dependence is associated with a large Wilson ratio, discussed below.

Specific heat measurements (Fig. 5) support the conclusion that the density of states in U$_3$Bi$_4$Ni$_3$ is gapped. A fit of the low temperature data to $C/T = \gamma_0 + \beta T^2$ gives a Sommerfeld coefficient $\gamma_0$ indistinguishable from 0 within experimental error for U$_3$Bi$_4$Ni$_3$. For U$_3$Bi$_4$Rh$_3$ in the absence of an applied magnetic field, a fit of $C/T$ above 6 K to the usual relation $C/T = \gamma_0 + \beta T^2$ gives $\gamma_0 = 117$ mJ/mol-U K$^2$ and $\beta = 1.5$ mJ/mol-U K$^4$ (red solid line). Taking this value of $\gamma_0$ and $\chi(2K) = 0.113$ emu/U-mol, we estimate a value for the Sommerfeld-Wilson ratio $R_W = \dfrac{\pi^2 k_B^2}{p_{eff}^2}\left(\dfrac{\chi}{\gamma}\right) = 18$, which is much larger than 2, typically found for heavy fermion systems, but more characteristic of nearly ferromagnetic metals or alloys, such as Pd ($R_w$=6-8), TiBe$_2$ ($R_w$=12), Ni$_3$Ga ($R_w$=40) [9]. The measured $C/T$ at lowest temperatures is larger than 117 mJ/mol-U K$^2$, eg., a simple extrapolation of $C/T$ from 0.4 K to T = 0 K gives a lower limit of ~ 200 mJ/mol-U K$^2$. Even using this value, $R_w$ is nearly 11. This large Wilson ration implies that U$_3$Bi$_4$Rh$_3$ is near a ferromagnetic instability, but there is no evidence for any long range order above 100 mK.

Below about 3 K, $C/T$ of U$_3$Bi$_4$Rh$_3$ follows a distinctly non-Fermi liquid temperature dependence. As shown in the Fig. 6, a good fit of the data (black solid line) over more than one decade in temperature is obtained using $C/T = -A \ln(T/T_0) + \beta T^2$, with A = 29.7 mJ/mol-U K$^2$, $T_0$ = 262 K and $\beta$ = 1.7 mJ/mol-U K$^4$. In the absence of more than trace amounts of second phase (RhBi, URh$_3$) in the x-ray pattern of U$_3$Bi$_4$Rh$_3$, it is unlikely that the upturn in $C/T$ below ~ 3K originates from impurities. Further, a modest field suppresses the upturn, and $C/T$ assumes a Fermi-liquid $C/T$=constant behavior below a crossover temperature that increases with increasing field (Figure 6). The $-A \ln(T/T_0)$ dependence of $C/T$ and its evolution with field is reminiscent of quantum-critical behavior observed in strongly correlated electron metals, such as CeCu$_{5.9}$Au$_{0.1}$ [10] and YbRh$_2$(Si$_{0.95}$Ge$_{0.05}$)$_2$ [11]. In this comparison, it also is noteworthy that the Sommerfeld-Wilson ratio ($R_W$=17.5) and an exponent $n$ characterizing a power-law divergence $\chi \propto T^{-\alpha}$ ($\alpha$ = 0.6) of YbRh$_2$(Si$_{0.95}$Ge$_{0.05}$)$_2$ [12] are comparable to that estimated for U$_3$Bi$_4$Rh$_3$.

Support for the possibility that U$_3$Bi$_4$Rh$_3$ might be near a quantum-phase transition is provided by resistivity measurements as a function of field. The inset of the lower panel in Fig. 7 shows the temperature dependence of representative resistivity curves after subtracting a residual value $\rho_0$, which was obtained by fitting $\rho(T)=\rho_0+A'T^n$ and letting $\rho_0$, $A'$ and $n$ be free parameters. As shown in this inset and summarized in the upper panel of Fig. 7, the exponent $n$ systematically increases from $n$=4/3 at zero field to $n$=2 for $\mu_0 H \geq$ 3T. The increase and saturation of $n$ with field is accompanied by a strong decrease and saturation, also for $\mu_0 H \geq$ 3T, of the coefficient $A'$, an evolution consistent with tuning the system from a non-Fermi-liquid to Fermi-liquid state. At a T=0 K ferromagnetic instability in an itinerant 3-dimensional system, theory predicts that $C/T$ should diverge as $-\ln T$ and, depending on the particular model of quantum criticality, that $(\rho(T)-\rho_0)$ should increase as $T^n$, where $n$=4/3 (Moriya), 5/3 (Lonzarich) or 1 (Hertz/Millis) [13]. With the large Wilson ratio for $U_3Bi_4Rh_3$ suggesting proximity to a ferromagnetic instability of the Fermi surface, the log-divergence in C/T, and power laws in resistivity, it appears that $U_3Bi_4Rh_3$ may be near a ferromagnetic quantum-critical point. The low-temperature magnetic susceptibility, however, is inconsistent with quantum criticality of itinerant ferromagnetism. In this case, these models predict $\chi \propto T^{-\alpha}$, with $\alpha = 4/3$ (Moriya and Lonzarich)[13] and not $\alpha = 3/4$ that we find in the same temperature range where specific heat and resistivity do agree with model predictions. On the other hand, the idea of local quantum criticality, which is argued to be relevant to $YbRh_2(Si_{0.95}Ge_{0.05})_2$ [14] does give an exponent in reasonable agreement with our observation. Reconciliation of these discrepancies remains an open question.

## 4. Discussion and Conclusions

We have succeeded in synthesizing $U_3Bi_4M_3$, where M = Rh, Ni, which are the first examples of a U-Bi-M 3-4-3 family. Within the larger family of U-based 3-4-3 compounds, electron count is an important factor that governs general trends in the nature of their ground states, and these trends also are found in our these materials. For example, $U_3Sb_4Ni_3$, $U_3Sb_4Pd_3$ and $U_3Sb_4Pt_3$ have nominally the same electron count as $U_3Bi_4Ni_3$ and all are non-metallic, even though their unit cell volumes differ by ~6%. Likewise, nominally isoelectronic $U_3Sb_4Co_3$, $U_3Sb_4Rh_3$ and $U_3Bi_4Rh_3$ are ferromagnetic, spin-glass like and nearly ferromagnetic metals, respectively. Though general trends are set by electron count, details are influenced by a volume-dependent hybridization between the 5f and ligand electrons. This is most apparent in the series that includes $U_3Bi_4Rh_3$. From entries in Table 2, ferromagnetic order at 10 K appears in the smallest cell-volume material $U_3Sb_4Co_3$; increasing the cell volume to $U_3Sb_4Rh_3$ produces glassy-like behavior from a competition between ferromagnetic tendencies of $U_3Sb_4Co_3$ and antiferromagnetic tendencies reflected in the large negative Weiss temperature of $U_3Sb_4Rh_3$; and finally, there is no order or glassiness in $U_3Bi_4Rh_3$, which has the largest cell volume, a large Sommerfeld-Wilson ratio and a large, negative Weiss temperature.

Additional experiments, such as neutron scattering, are needed to establish more definitively the origin of non-metallic behavior and the weakly temperature-dependent magnetic susceptibility of $U_3Bi_4Ni_3$. We have suggested that these behaviors may arise from hybridization of 5f and ligand electrons, analogous to what is found in $Ce_3Bi_4Pt_3$, but we can not rule out a simple band-structure interpretation. On the other hand, the large Sommerfeld-Wilson ratio, a logarithmic dependence of C/T, $\rho \propto T^n$, where n < 2 in zero field, and the evolution of these to Fermi-liquid behaviors for $\mu_0 H \geq 3T$ strongly suggest that $U_3Bi_4Rh_3$ is near a ferromagnetic quantum-critical point. Given the trends in the isoelectronic 3-4-3 series with $U_3Bi_4Rh_3$, we would anticipate that applying pressure to $U_3Bi_4Rh_3$ should induce long-ranged ferromagnetic order within an accessible, albeit high, pressure range needed to reduce its cell volume by ~15%.

**Acknowledgements**

Work at Los Alamos and Princeton was performed under the auspices of the US Department of Energy, Office of Science.

**Table 1**

Structural parameters for $U_3Bi_3Ni_3$. Space group I -4 3 d. Crystallographic sites: U: 12a (3/8,0,1/4); Bi: 16c (x,x,x); Ni: 12b (7/8,0,1/4). A flat plate surface roughness correction to account for sample absorption was applied. For $U_3Bi_4Rh_3$, a preferred orientation correction was also used in the final refinement.

|  | a (Å) | $x_{Bi}$ | $U_U$ | $U_{Bi}$ | $U_{Ni/Rh}$ |
|---|---|---|---|---|---|
| **$U_3Bi_4Ni_3$** | 9.5793(1) | 0.0829(2) | 0.0136(7) | 0.016(3) | 0.0162(8) |
| **$U_3Bi_4Rh_3$** | 9.7273(1) | 0.0884(1) | 0.0081(7) | 0.0077(5) | 0.0028(9) |

Statistics $U_3Bi_4Ni_3$: $\chi^2 = 1.285$, $R_{wp} = 12.56\%$, $R_p = 9.84\%$

Statistics $U_3Bi_4Rh_3$: $\chi^2 = 0.5579$, $R_{wp} = 18.21\%$, $R_p = 13.53\%$

**Table 2**

Physical properties for selected Uranium 3-4-3 compounds.

| Compound | a (Å) | $\Theta_{CW}$ (K) | $\mu_{eff.}$ ($\mu_B$/U-mol) | $\chi$ (4.2K) ($10^{-3}$ emu / U-mol) | $\rho$(300K) (m$\Omega$ cm) | $\gamma_0$ (mJ/K$^2$ U-mol) |
|---|---|---|---|---|---|---|
| U$_3$Bi$_4$Ni$_3$ | 9.5800(9) | -117 | 3.6 | 7.3 | 5.4 | ~ 0 |
| U$_3$Sb$_4$Ni$_3$ [6] | 9.393 | -99 | 3.65 | 11 | 2930 | 2 |
| U$_3$Bi$_4$Rh$_3$ | 9.7289(5) | -180 | 3.4 | 63 | 0.48 | 200* |
| U$_3$Sb$_4$Rh$_3$ | 9.501(1) | -110 | 3.2 | --- | 0.62 | --- |
| U$_3$Sb$_4$Co$_3$ | 9.284 | +11.7 | 2.1 | --- | ~0.43 | --- |

* at 0.4K

**Figure Captions**

**Figure 1.** (Color online) Observed (blue circles) and calculated (solid red line) x-ray diffraction patterns for $U_3Bi_4Ni_3$ at room temperature. The difference plot is shown at the bottom and vertical bars represent the Bragg peak positions for $U_3Bi_4Ni_3$ (upper set) and Bi (lower set).

**Figure 2.** (Color online) Temperature dependence of resistivity ρ(T) for $U_3Bi_4Ni_3$ (upper panel) and for $U_3Bi_4Rh_3$ (lower panel). The upper inset compares resistivity (log scale) of $U_3Bi_4Ni_3$ and $Th_3Bi_4Ni_3$, with non-metallic behavior visible for both compounds. The inset of the lower panel shows the low- temperature resistivity of $U_3Bi_4Rh_3$ under magnetic fields of 0 and 1T. The slight drop in ρ(T) below 2K ($\mu_0H=0T$) is due to the presence of a tiny amount of superconducting RhBi on the crystal surface of $U_3Bi_4Rh_3$. Applying a field ($\mu_0H=1T$) suppresses that superconductivity and ρ(T) can be fitted between 0.5 K and 8 K by $\rho(T)=0.163+(7.4*10^{-4})*T^{1.69}$.

**Figure 3.** (Color online) Near – Fermi level photoemission data. Symmetrization of the density of states with respect to Fermi level was performed for gap size estimation in $U_3Bi_4Ni_3$. No gap was observed in $U_3Bi_4Rh_3$.

**Figure 4.** (Color online) DC magnetic susceptibility χ vs temperature, at the applied field of $\mu_0H=0.1T$, for $U_3Bi_4Rh_3$ (solid blue circles) and for $U_3Bi_4Ni_3$ (open black circles). The inset shows the magnetic susceptibility vs $T^{-3/4}$ for $U_3Bi_4Rh_3$.

**Figure 5.** (Color online) Specific heat divided by temperature ($C/T$) as a function of temperature for both $U_3Bi_4Rh_3$ (open blue circles) and $U_3Bi_4Ni_3$ (open squares). The blue line represents a very good fit to $C/T = -A \ln(T/T_0) + \beta T^2$, and the red line corresponds to a fit to $C/T = \gamma + \beta T^2$ for T > 8 K and its extrapolation to lower temperatures. See text for details.

**Figure 6.** (Color online) Electronic contribution to the specific heat divided by temperature as a function of temperature on a logarithmic scale for $U_3Bi_4Rh_3$. A good fit of the experimental data (black solid line) is obtained using $C/T = -A \ln(T/T_0) + \beta T^2$. Applying a magnetic field supresses the C/T upturn, and Fermi-liquid behavior ($C/T$ = const.) is recovered.

**Figure 7.** (Color online) Field dependence of the exponent $n$ (upper panel) and coefficient $A'$ (lower panel) obtained from fitting $\rho(T) = \rho_0 + A'T^n$. The inset displays a log-log plot of ($\rho(T) - \rho_0$) vs temperature at varying magnetic fields (0.5, 1 and 9T)

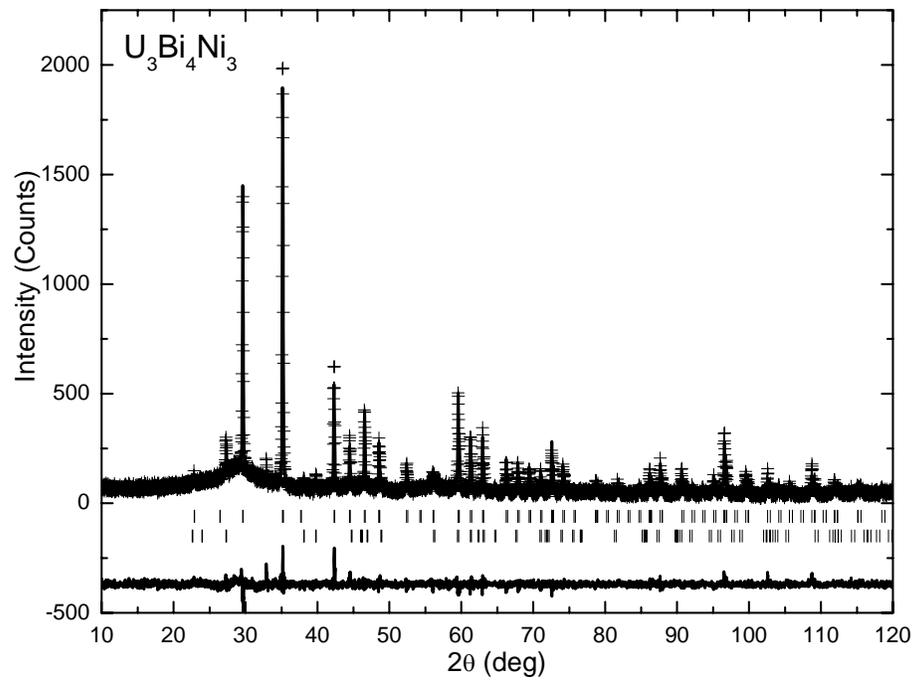

**Fig. 1**

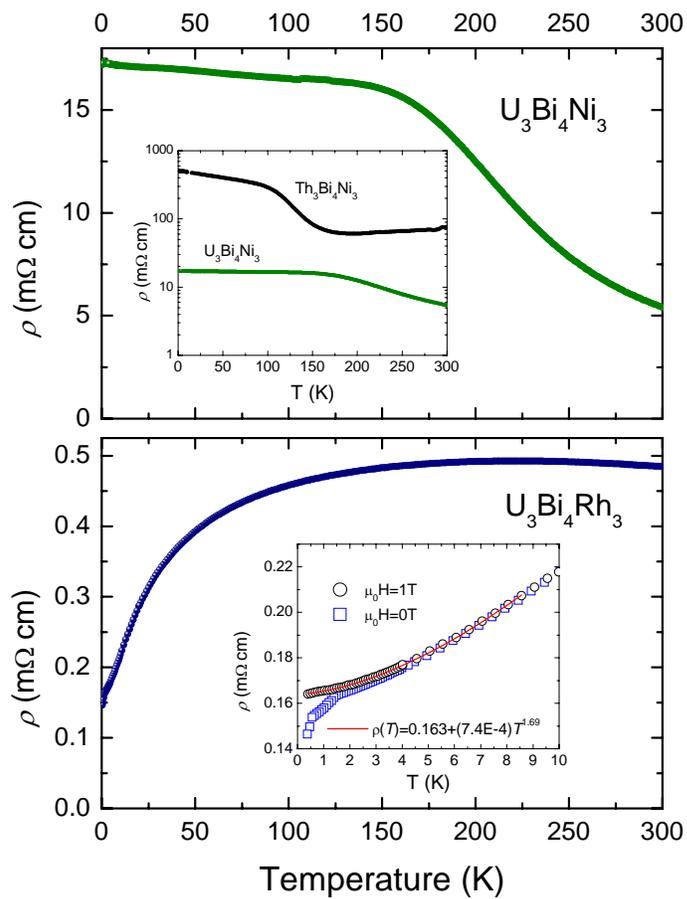

**Fig. 2**

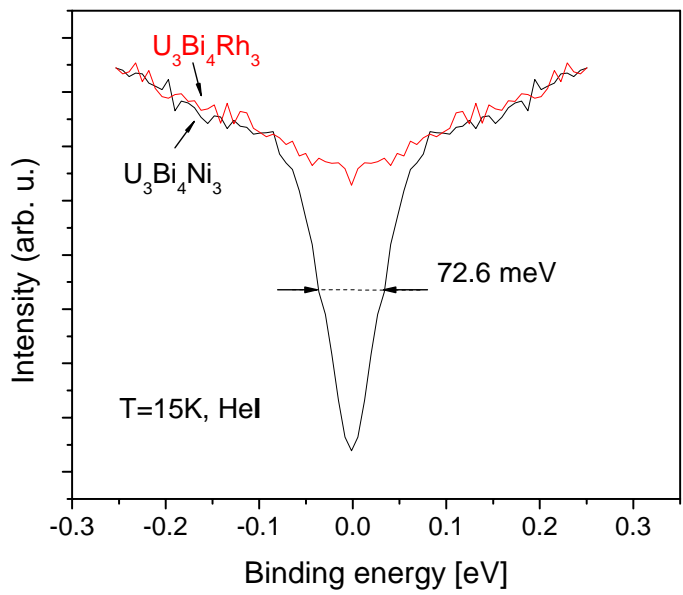

**Fig. 3**

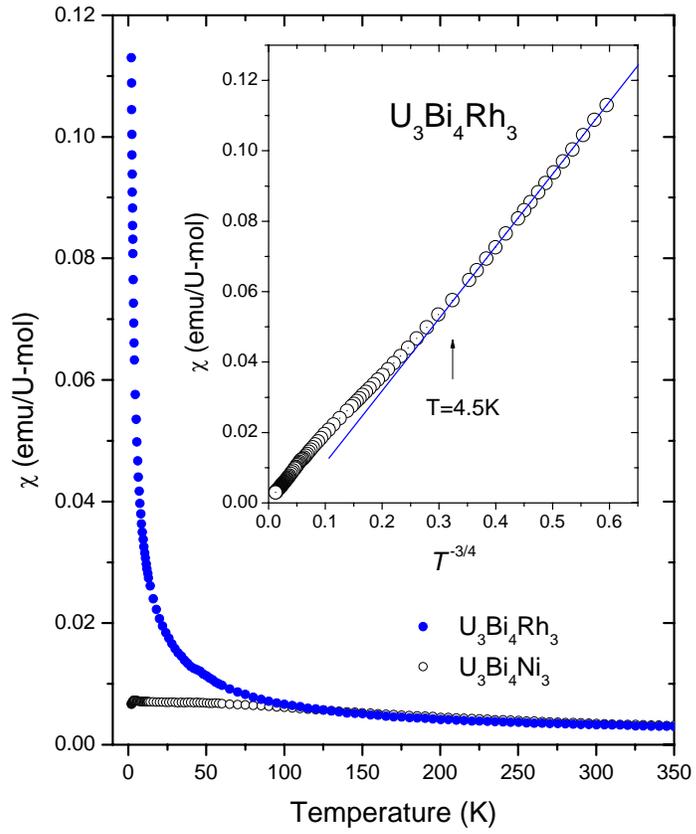

**Fig. 4**

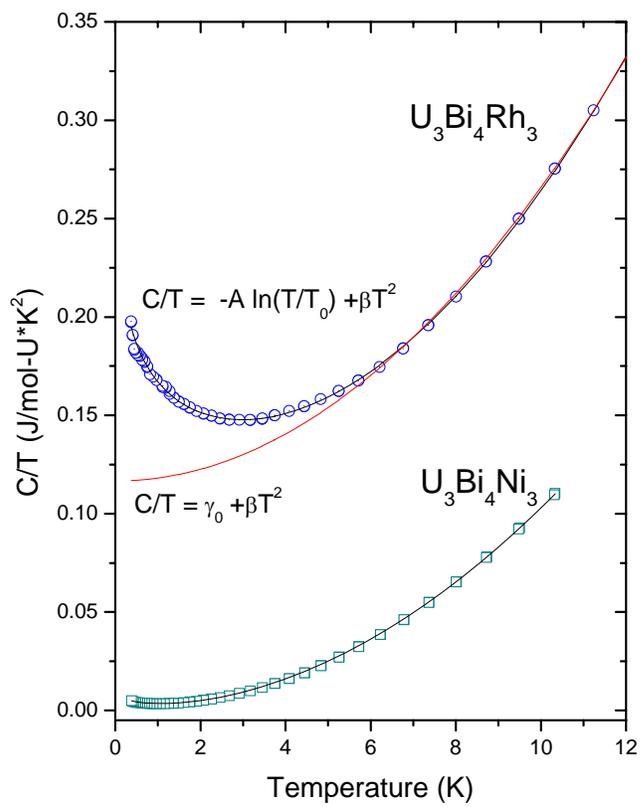

**Fig. 5**

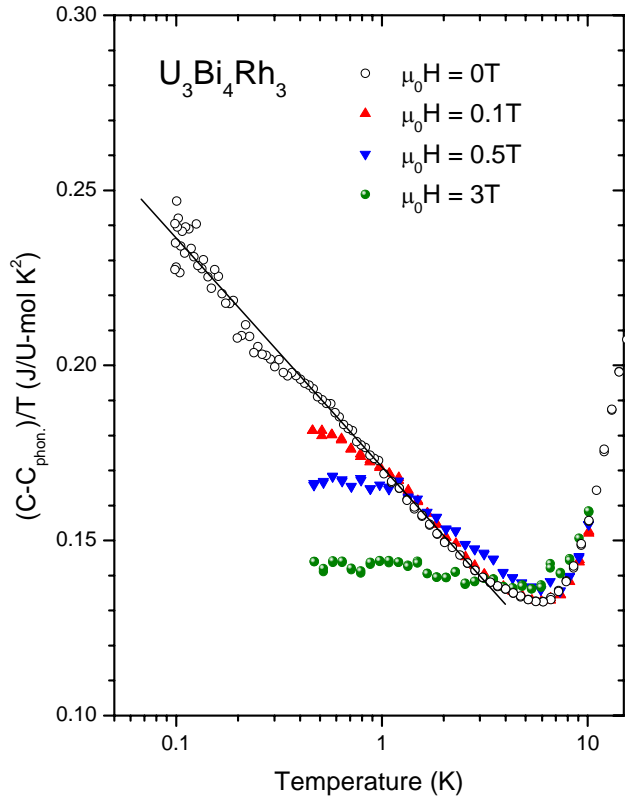

**Fig. 6**

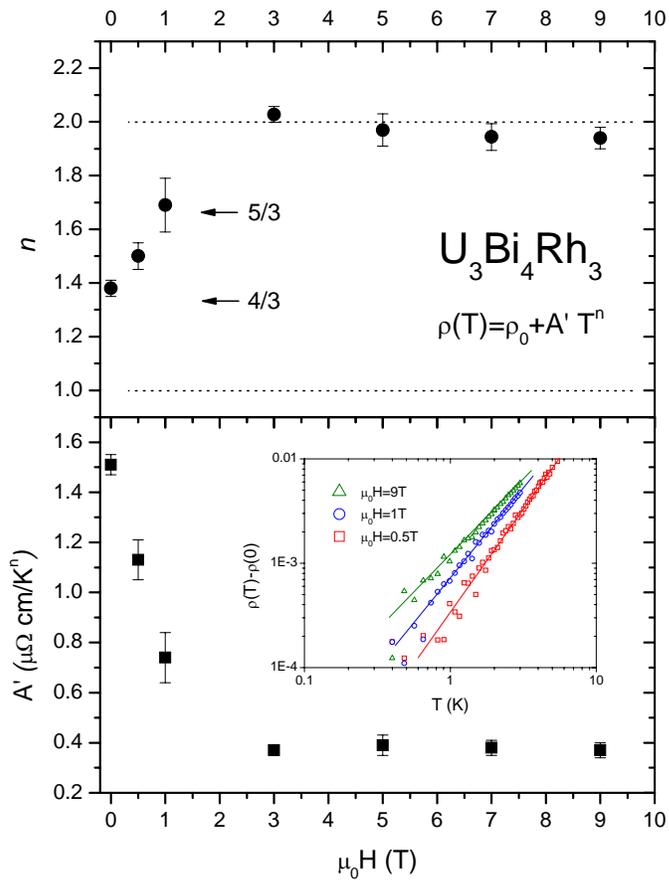

**Fig. 7**

**References**

[1] M.F. Hundley; P.C. Canfield, J.D. Thompson, Z. Fisk, J.M. Lawrence, Phys. Rev. **42**, 6842 (1990)

[2] S. V. Vonsovsky, Yu. A. Izumov, and E. Z. Kuramev, *Superconductivity in Transition Metals*, Springer-Verlag, Berlin, 1982

[3] T. Endstra, G.J. Nieuwenhuys, J.A. Mydosh and K.H.J. Buschow, J. Magn.Magn. Mat. **89**, L273 (1990)

[4] R. Ferro, Atti della Accademia Nazionale dei Lincei, Classe di Scienze Fisiche, Matematiche e Naturali, Rendiconti, **13,** 53 (1952)

[5] A. E. Dwight, Acta Crystall. B **33**, 1579 (1977)

[6] T. Takabatake, H. Fujii, S. Miyata, H. Kawanaka, Y. Aoki, T. Suzuki, T. Fujita, Y. Yamaguchi, J. Sakurai, J. of Phys. Soc. Japan **59**, 4412 (1990)

[7] V.H. Tran, Z. Bukowski, J. Stepien-Damm, A.J. Zaleski, D. Badurski, R. Gorzelniak, C. Sulkowski, R. Troc, J. of Phys.: Cond. Matt. **17**, 3597 (2005)

[8] A. Severing, J. D. Thompson, P. C. Canfield, Z. Fisk, P. Riseborough, Phys. Rev. B **44**, 6832 (1991)

[9] S. R. Julian et al., Physica (Amsterdam) **259B–261B**, 928 (1999), and references therein.

[10] H. Von Lohneysen, T. Pietrus, G. Portisch, H.G. Schlager, A. Schröder, M. Sieck, T. Trappmann, Phys. Rev. Lett. **72**, 3262 (1994)

[11] J. Custers, P. Gegenwart, H. Wilhelm, K. Neumaier, Y. Tokiwa, O. Trovarelli, C. Geibel, F. Steglich, C. Pépin & P. Coleman, Nature **424**, 524 (2003)

[12] P. Gegenwart, J. Custers, Y. Tokiwa, C. Geibel, and F. Steglich, Phys. Rev. Lett. **94**, 076402 (2005)

[13] G.R. Stewart, Rev. Mod. Physics **73**, 797 (2001)

[14] P. Gegenwart, J. Custers, Y. Tokiwa, C. Geibel, and F. Steglich, Phys. Rev. Lett. **94**, 076402 (2005)